\documentclass[draft]{llncs}
\usepackage{ucs}
\usepackage[utf8x]{inputenc}
\usepackage[T1]{fontenc}
\usepackage[english,ngerman]{babel}
\usepackage{graphicx}
\usepackage{amssymb}
\usepackage{amsmath}

\begin{document}
\selectlanguage{english}
\title{Data}
\author{2018-06-07, Johannes Reich\inst{1}}
\institute{\email{johannes.reich@sap.com, SAP SE}}


\maketitle
\begin{abstract}
The contribution of this article is a data concept that is essentially based on the two concepts of information and computable functionality. In short, data is viewed as typed information.

A data type is defined as a pair of a set of distinguishable characters (an alphabet) and a set of operations (computable functions) that operate on this alphabet as domain.

Two different ways of subtyping in the sense of Liskov and Wing are described, one for restriction and one for extension of existing types. They lead to two different partial orders on types.

It is argued that the proposed data concept matches the concept of characteristics (Merkmale) of the automation industry. 
\end{abstract}
\keywords{
data,
information,
data type,
operation type,
type system,
type relation,
subtyping,
computability,
characteristics,
interoperability
}

%
\section{Introduction}
%
What are data? Or --- what is data? What is the difference between information and data? It might seem strange that in 2018 someone writes an article about the concept of data. But, one of the consequences of the youth of informatics, in contrast to other, more settled disciplines, like mathematics or physics, seems to be the heterogeneity of even some of its rather fundamental concepts - like data or information. 

Surely, there will not be the one-and-only meaning of the term ''data'' in our natural language. But it seems to be a worthwhile undertaking to develop a mutually agreed meaning in the specialist language of the informatics people.
  
The Merriam-Webster Dictionary\footnote{https://www.merriam-webster.com/dictionary/data} says that ''data'' is used both as a plural noun (like earnings) and as an abstract mass noun (like information). It gives three different definitions of data, all based on the notion of information and two also explicitly referring to their processing:
\begin{enumerate}
\item Factual information (such as measurements or statistics) used as a basis for reasoning, discussion, or calculation.
\item Information output by a sensing device or organ that includes both useful and irrelevant or redundant information and must be processed to be meaningful.
\item Information in numerical form that can be digitally transmitted or processed.
\end{enumerate}

The data concept in this sense also dominates the very influential entity-relationship-model of Peter P.-S. Chen \cite{Chen1976_Entity} and others, the de-facto standard for data models. There entities (individual, identifiable objects of the real world) are characterized by attributes and relationships. It was extended by generalization and specialization \cite{Smith1977_Database}. 

However, a substantial part of the scientific community has a different model in mind when reflecting about data and information. Chaim Zinn \cite{Zins2007_Conceptual} documented 130 definitions of data, information, and, in addition, knowledge from 45 scholars of 16 countries.  Many scholars seemed to be the opinion that knowledge can be defined in terms of information and information can be defined in terms of data, following a model sometimes called the ''Knowledge  Pyramid'' (e.g. \cite{Rowley2007_Wisdom}).

This is surprising as it was the notion of information as developed by Ralph V. L. Hartley \cite{Hartley1928}, Claude Shannon \cite{Shannon1948}, and others that stood at the beginning of the field of informatics. It was their breakthrough idea to introduce a completely new perspective on the physical world that disregards the quality of the physical states, be it voltage, pressure, current, etc. and takes interest only in the values of these quantities as they can be distinguished as values of ''information''. 

Thereby communication became amenable to quantification and with communication, transport and processing of information were separated. Information becomes transported and is locally processed. By identifying ''processing of information'' with ''attributing meaning to information'' we can say (tautologically) that the ''meaning of information is attributed by processing''. Then we can qualify any concept that classifies the processing of information as a semantic concept.  

The contribution of this article is a semantic concept in this sense as we combine the two concepts of information and computable functionality with typing. Types were introduced to informatics by Alonso Church in 1940 \cite{Church1940} as a means to guarantee well-formedness of formulas of his $\lambda$-calculus\cite{Church1932}. Beside Turing machines and the theory of computable functions\footnote{or ''recursive functions'' as they were called.}, this calculus is one of the models of computation. In the typed $\lambda$-calculus, simple types $\sigma$ for simple terms and function types $\sigma \rightarrow \tau$ for $\lambda$-terms are defined. Church did not commit himself to any concrete interpretation, but pointed out that ``We purposely refrain from making more definite the nature of the types ..., the formal theory admitting of a variety of interpretations in this regard''. 

Indeed, typing in informatics is usually tied to a formal calculus of computation, for example when Luca Cardelli says, “the fundamental purpose of a type system is to prevent the occurrence of execution errors during the running of a program.” \cite{Cardelli1997_TypeSystems}. Accordingly, type systems are usually viewed as a ''syntactic method for proving the absence of certain program behaviors by classifying phrases according to the kinds of values they compute'' \cite{Pierce2002}.

But our approach rests on the theory of computable functions as it was developed by Kurt Gödel, Stephen Kleene \cite{Kleene1936} and others and therefore does not presuppose any concept of a formal programming calculus. The main purpose of the presented approach is to introduce a useful, mathematically founded data concept that captures somehow most of the scope of the intuitive meaning of this concept and, in addition to that, allows the derivation of further useful consequences. One such consequence is surely to use our knowledge about data types in our design of programming languages for the important purpose Luca Cardelli points out.

According to their semantic character, informatical data types are supposed to carry quite desirable properties. They ought to assign meaning to bits and bytes, they seem to carry the intent of the programmer and last but not least prevent inconsistencies in data processing. Even prominent institutions as the UN have taken serious effort to overcome semantic issues in business communication with the help of a data type system \cite{UMM2003UserGuide,CCTS2009}. 


\subsection{Preliminaries}
Elements and functions are denoted by small letters, sets and relations by large letters, and mathematical structures by large calligraphic letters. 
The components of a structure may be denoted by the structure's symbol or, in case of enumerated structures, index as subscript. The subscript is dropped if it is clear to which structure a component belongs.

To talk about information transport and processing, we have to agree on the names of these distinguishable values. We name these values ''characters''. Thus, a character can be distinguished from other characters and has no other further properties. We name enumerable sets of characters ''alphabets''. If not stated otherwise, characters can be vectors. 

%
\section{Data}
%
Let us assume, that our informatics perspective has already resulted in a set of alphabets $V=\{V_1, \dots, V_m\}$ , signals (as a mapping from a time domain $T$ onto the value set $V$) can represent. 

\subsection{Computable functionality with natural numbers}
As the denotation of the distinguishable values of the information sets are arbitrary, looking at them as natural numbers, as the pioneers of computability did, is possible.  

Be $F_n$ the set of all functions on natural numbers with arity $n$ and there exists a set of elementary computable functions (the successor, the constant and the identity function). Then, based on work of Kurt Gödel, Stephen Kleene \cite{Kleene1936} showed that there are three rules to create all computable functions:
\begin{enumerate}  
\item {\bf\it Comp:} Be $g_1, \dots, g_n \in F_m$ computable and $h\in F_n$ computable, then $f = h(g_1, \dots, g_n)$ is computable.\label{computation_1st_rule}
\item {\bf\it PrimRec:} Are $g\in F_n$ and $h\in F_{n+2}$ both computable and $a\in \mathbb{N}^n$, $b\in \mathbb{N}$ then also the function $f\in F_{n+1}$ given by $f(a, 0) = g(a)$ and $f(a, b+1) = h(a, b, f(a,b))$ \label{computation_2nd_rule} is computable.
\item {\bf\it $\mu$-Rec:} Be $g\in F_{n+1}$ computable and $\forall a\exists b$ such that $g(a,b) = 0$ and the $\mu$-function $\mu_b[g(a,b) = 0]$ is defined as the smallest $b$ with $g(a,b) = 0$. Then $f(a) = \mu_b[g(a,b) = 0]$ is computable.\label{computation_3rd_rule}
\end{enumerate}

\subsection{Computable functionality with arbitrary alphabets \label{ss_comp_func_arb_alph}}
We now reformulate the three computation rules for arbitrary alphabets:
\begin{definition} \label{def_comuputation_arb_alph} We call a computable function with alphabets $V$ as domain and $W$ as codoomain an operation.
Be $V=\{V_1, \dots, V_m\}$ a set of alphabets and $E=\{e_1, \dots, e_n\}$ a set of elementary operations with $e_j:V_{i_1}\times \dots \times V_{i_{k_j}} \rightarrow V_{l_j}$. To proceed we pick three appropriate alphabets $X, Y, Z$\footnote{We refrained from rephrasing also the enumerating loop parameters as we would then have to introduce a successor relation on these alphabets, which would bring us either way back to the natural numbers.}.

\begin{enumerate}
\item {\bf\it Comp:} Be $g_i:X_i\rightarrow Y_i$, $(i=1\dots n)$ with $Y = Y_1\times \dots \times Y_n$ and $h:Y\rightarrow Z$ both computable, then $f = h(g_1, \dots, g_n)$ is computable. \label{computation_1st_rule_b}
\item {\bf\it PrimRec:} Are $g:X\rightarrow Y$ and $h:X\times \mathbb{N}\times Y \rightarrow Y$ both computable and $a\in X$, $b\in \mathbb{N}$, then also the function $f:X\times \mathbb{N}\rightarrow Y$ given by $f(a, 0) = g(a)$ and $f(a, b+1) = h(a, b, f(a,b))$ is computable. \label{computation_2nd_rule_b}
\item {\bf\it $\mu$-Rec:} Be $g:X\times \mathbb{N} \rightarrow \mathbb{N}$ computable and $\forall a\in X \exists b\in \mathbb{N}$ such that $g(a,b) = 0$ and the $\mu$-operation $\mu_b[g(a,b) = 0]$ is defined as the smallest $b$ with $g(a,b) = 0$. Then $f(a) = \mu_b[g(a,b) = 0]$ is computable.\label{computation_3rd_rule_b}
\end{enumerate} 

We name the set of all operations $O$ derivable from the set of alphabets $V$ and the set of elementary operations $E$ with the computation rules the closure of $E$ with respect to $V$ and write $O = closure_V(E)$. 
\end{definition}

\subsubsection{Currying}
An operation $f:V_{i_1}\times \dots \times V_{i_k} \rightarrow V_{j} $ depends on $k$ variables $x_1, \dots, x_k\in V_{i_1}\times \dots \times V_{i_k}$. 
As we want to focus on operations on a single variable $x_l\in V_{i_l}$ with $1\leq l \leq k$, we need a procedure, to transform a function with multiple arguments into a sequence of functions, each with a single argument. This is well known from functional programming and was named ''Currying'' by Christopher Strachey in 1967 in honor of the logician Haskell Curry. 

\begin{definition}
Given an operation $f:V_{i_1}\times \dots \times V_{i_k} \rightarrow V_{j}$, depending on at least 2 variables, that is $k\geq 2$, and an element $a\in V_{i_l}$ with $i_l \in \{i_1, \dots, i_k\}$ then $h(\alpha) = f|_{x_l = \alpha}$ is the restriction of $f$ on the value $x_l = \alpha$. The interpretation of $h$ as a mapping from $V_{i_l}$ to the set of functions mapping $V_{i_1}\times\dots\times V_{i_{l-1}}\times V_{i_{l+1}}\times\dots\times V_{i_k}$ to ${V_j}$ is the desired function with domain $V_{i_l}$ and we write $h = curry_l(f)$. For $k=1$ we define $ curry_1(f) = f$
\end{definition}

Please note, that although $f$ and $h(\alpha)$ are operations in the sense of Def. \ref{def_comuputation_arb_alph}, $h = curry_l(f)$ is generally not, because its codomain is not an alphabet in the given sense of a set of characters, but a set of operations. We call such a function a curried-operation, or ''curriedop''. Only curriedops which result from $curry_1$ are ordinary operations.

If we want to express currying with respect to a specific domain set $X$ we also write $H = curry_X(f)$ where $H$ is the set of all functions where the corresponding variables represent elements of $X$. 

\subsection{Data types}\label{ss_data_types}
If we say that a character is a datum we say two things: first, this character belongs to a certain alphabet. Second, all of the characters of this alphabet can be processed by all of the operations of a certain set. 
Actually, it is generally agreed that a data type defines a set of (data) values together with a set of operations, having this value set as their domain  \cite{Sebesta2007,vanRoyHaridi2004,WattFindlay2004,Mitchell2000,Pierce2002}. Although, there had been other opinions viewing types only as sets of values (e.g. \cite{CardelliWegner1985}) or as equivalence classes of variables (e.g. \cite{ParnasShoreWeiss1976}).
So, we define:

\begin{definition}
Be $W=\{W_1, \dots, W_m\}$ a set of alphabets, $E=\{e_1, \dots, e_n\}$ a set of elementary operations on $W$ and $O=closure_W(E)$. A data type ${\cal T}$ is a pair of two nonempty sets ${\cal T}=(V, F)$, an alphabet $V\in W$ and the set $F$ of all curriedops with $V$ as their domain, that is $F=curry_V(O)$. We then say that a character $c\in V$ is of type ${\cal T}$ and call it a datum. We call the set $T = \{{\mathcal T}_1, \dots, {\mathcal T_m}\}$ with $V_{{\mathcal T}_i} = W_i$ the type system with respect to its base $(W, E)$.
\end{definition}

\subsection{Data type composition}
We can compose new types from existing types in the following product sense:
\begin{definition}
Be $T = \{{\mathcal T}_1, \dots, {\mathcal T}_n\}$ a type system with base $(W,E)$. Then we can construct a product type ${\mathcal T}^*$ with $V^* = W_{k_1} \times \dots \times W_{k^*}$ and\\ $F^* = curry_{V^*}(closure_{W\cup V^*}(comp^*(closure_W(E))))$, where the composition operator $comp^*$ provides the necessary elementary operations in the sense of Def. \ref{def_comuputation_arb_alph} for the new type.
\end{definition}

A simple example would be the type system $\{Real\}$ with the base $(\mathbb{R}, \{+,-\})$ that is extended to the type ${\mathcal C}$ by defining $V_{\mathcal C} = \mathbb{R} \times \mathbb{R}$ and the composition operator provides three operations $\{create_{\mathcal C}, +_{\mathcal C}, *_{\mathcal C}\}$ where $create_{\mathcal C}: \mathbb{R} \times \mathbb{R} \rightarrow V_{\mathcal C}$ with $create_{\mathcal C}(x,y) = (x,y)$, $+_{\mathcal C}: V_{\mathcal C} \times V_{\mathcal C} \rightarrow V_{\mathcal C}$ with $+_{\mathcal C}(x,y) = (x_1 + y_1, x_2 + y_2)$ and $*_{\mathcal C}: V_{\mathcal C} \times V_{\mathcal C} \rightarrow V_{\mathcal C}$ with $*_{\mathcal C}(x,y) = (x_1*y_1-x_2*y_2, x_1*y_2+x_2*y_1)$.

\subsection{Data type relations}
A key concept of typing is to derive new types from already defined ones by not only relating the alphabets, but also the set of operations. Barbara H. Liskov and Jeannette M. Wings \cite{LiskovWing1994} formulated the ''Substitutional Principle'' of subtyping: Let $\phi(x)$ be a property of all objects $x$ of type $T$. Then $\phi(y)$ should be true for objects $y$ of type $S$ where $S$ is a subtype of $T$. 

Other authors seem to assume that subtyping means subset relations between the value sets while leaving the set of operations invariant (e.g. David A. Watt and William Findlay in \cite{WattFindlay2004}, p.191 or Benjamin C. Pierce \cite{Pierce2002}, p.182) while other authors (e.g. John C. Mitchell \cite{Mitchell2000}, p. 704) relate subtyping to a subset relations between the set of operations.  

We will see that there are at least two Liskov-Wing-subtype relations for data types creating two different partial orders on our type-graph. To proceed, we need the following relation between two sets of operations.

\begin{definition}
Be $F^V, F^{V'}$ two sets of curriedops with the domains $V \supseteq V' \neq\emptyset$. If $F^{V'}$ contains all restricted curriedops $f' = f|_{V'}$ with $f\in F^V$ in addition to all curriedops operating only on the alphabet $V'$, we say that $F^{V'}$ contains restricted $F^V$ and write $F^V \sqsubseteq F^{V'}$.
\end{definition}

\subsubsection{Restriction/Expansion} 
The first Liskov-Wing-subtype property we look at is $\Phi(x)=$''Character x is being processable by every operation $f\in F$ of type ${\cal T}$''. It is useful for restricting an existing type.

\begin{definition} Be ${\cal T} = (V, F)$ a defined data type. We derive a restricted type ${\cal T}'= (V', F')$ by requiring $V \supseteq V' \neq \emptyset$ and $F \sqsubseteq F'$. We call ${\cal T}$ the expanded type and ${\cal T}'$ the restricted type.
\end{definition}

From the subset relations of the alphabets immediately follows:
\begin{lemma}
Every character $c\in V'$ of the restricted data type ${\cal T}'$ can be processed by every curriedop $f\in F$ of the expanded type ${\cal T}$. 
\end{lemma}

In other words, every character of type ${\cal T}'$ can be treated as if it were of type ${\cal T}$. We also say that every character of type ${\cal T}'$ can be ''safely R-casted\footnote{''R'' stands for restriction as the basic subtyping mechanism.} (=expanded)'' to type ${\cal T}$. 
Clearly, the following subtyping proposition holds. 
\begin{proposition} Be ${\cal T}'$ a restricted data type of ${\cal T}$, then ${\cal T}'$ is an (R-)subtype of ${\cal T}$ in the Liskov-Wing sense with respect to the property $\Phi(x)=$''Character x is being processable by every operation $f\in F$ of type ${\cal T}$''.
\end{proposition}

Example: Be $Char$ the type with all printable characters as alphabet. It is possible to define the R-subtype $Alphanum$, relating to all alphanumeric characters, by restricting the alphabet in relation to the alphabet of $Char$. 
\begin{itemize}
\item $V_{Alphanum} \subseteq V_{Char}$: The set of alphanumeric characters is just a subset of the set of all possible printable characters.
\item $F_{Alphanum} \sqsupseteq F_{Char}$: Each curriedop capable of processing all elements of $V_{Char}$  is also able to process all elements of $V_{Alphanum}$.
\end{itemize}

Thus, a character of type $Alphanum$ can safely be R-casted (or expanded) to $Char$, but not vice versa.

\subsubsection{Truncation/Extension}
The second Liskov-Wing-subtype property we look at is $\Phi(x)=$''The projection $\pi(x)$ of character x is being processable by every operation $f\in F$ of type ${\cal T}$''. It is useful for extending an existing type.

\begin{definition} Be ${\cal T} = (V,F)$ a data type. We derive an extended type ${\cal T}' = (V', T')$ by requiring the existence of a projection function\footnote{A projection function $\pi$ fulfills the equality $\pi = \pi \circ \pi$. Therefore its codomain must be a subset of its domain.} $\pi:V\cup V' \rightarrow V$ such that $\pi(V) = V$, $V\supseteq \pi(V') \neq \emptyset$, and $F^V \sqsubseteq F^{\pi(V')}$. We call ${\cal T}$ the truncated type and ${\cal T}'$ the extended type.
\end{definition}

And again, from the subset relation $\pi(V') \subseteq V$ it follows immediately:
\begin{lemma} The projection $\pi(c)$ of every character $c\in V'$ of the extended data type ${\cal T}'$ can be processed by every curriedop $f\in F$ of the truncated data type ${\cal T}$.
\end{lemma}

In other words, every projected character of type ${\cal T}'$ can be treated as if it were of type ${\cal T}$. We also say that every character of type ${\cal T}'$ can be ''safely P-casted\footnote{''P'' stands for projection as the basic subtype mechanism.} (=truncated)'' to type ${\cal T}$. 
Finally, the following subtyping proposition holds. 
\begin{proposition} Be ${\cal T}$' an extended data type of ${\cal T}$ with the required projection $\pi$. Then ${\cal T}'$ is a (P-)subtype of ${\cal T}$ in the Liskov-Wing sense with respect to the property $\Phi(x)=$''The projection $\pi(x)$ of character x is being processable by every operation $f\in F$ of type ${\cal T}$''.
\end{proposition}

Example: Be $Alphanum20$ a type having the alphabet of all alphanumeric characters together with an extra character $unknown$ in sequences of length 20 as value set. Then we can construct a P-subtype $Char40$ as an extension by providing a projection $\pi:V_{Char40} \rightarrow V_{Alphanum20}$ such that $\pi(c_i) = c_i$ if the $i$-th character is alphanumeric and else $\pi(c_i) = unknown$. 

\begin{itemize} 
\item $V_{\pi(Char40)} \subseteq V_{Alphanum20}$: Each element in the projected set $V_{\pi(Char40)}$ is also part of the value set $V_{Alphanum20}$ of the truncated type.
\item $F_{\pi(Char40)} \sqsupseteq F_{Alphanum20}$: Each curriedop capable of processing all elements of $_{Alphanum20}$ is also able to process all elements of the projected set $\pi(Char40)$ of the extended type $Char40$.
\end{itemize}
 
With the truncation function being the projection, a character of type $Char40$ can be safely P-casted (or truncated) to $Alphanum20$, but not vice versa.

As the example illustrates, extension does not just mean to extend the value set, but also to assure that really all projected values belong to the original alphabet and therefore can be processed by the original curriedops. Additional dimensions of the extended type can simply be truncated. 

\subsection{Data type hierarchies}
Obviously, we now have two ways to create data type hierarchies: either by starting from some top level type and restrict it more and more, or by starting from some bottom-level type and extend it more and more. However, in both cases the subtypes are the derived types.
 
Both subtypings define a partial order on the their derived types. As both subtypings can be combined, we get a type graph with two kinds of edges. 

As long as extension is restricted to extend the elements of an alphabet without changing its dimensions, we can have circles in our type graph, resulting in safe casting in ''opposite'' directions. 

Example: Be ${\cal T}$ a type with $V = \{a\}$. We extend it to the P-subtype ${\cal T}^{ext}$ with $V^{ext} = V \cup \{b\}$ and $\pi:\{a, b\} \rightarrow \{a\}$ such that $\pi(a) = a$ and $\pi(b) = a$. We can now restrict ${\cal T}^{ext}$ to the R-subtype ${\cal T}'$ with $V' = {V^{ext} / \{b\}} = \{a\}$. Obviously ${\cal T} = {\cal T}'$. 

We can now cast a character of type ${\cal T} = {\cal T}'$ safely to ${\cal T}^{ext}$ and back. Any character of the original set {\cal V} (in our example only $a$) thereby remain invariant, but an eventually chosen character of type ${\cal T}^{ext}$ that is not an element of $V$ is changed by P-casting to some character in $V$ (in our example to $a$). 

If we extend a type's alphabet $V$ by adding some dimensions, then we have no circles anymore, because of the different requirement between the subset relation between the alphabets of type restriction, which requires a nonempty subset, and the projection relation between the alphabets of type extension, which allows dimension reduction.  

\subsection{Relation to characteristic based system description}
Standardizing the meaning of system properties by stipulating their types is a common technique (e.g. \cite{CCTS2009,IEC61360-1}). In automation engineering, there have been substantial efforts to standardize the meaning of characteristics (German ''Merkmale'') to simplify interoperability (see for example eClass, Prolist).

According to Ulrich Epple \cite{Epple2011_Merkmale,Epple2017_Properties}, a characteristic is a classifying property of a system whose manifestations can be represented by single values - which is essentially our definition of the alphabets of data types in section \ref{ss_data_types}. Hence, each characteristic in this sense can be assigned a type in our sense. 

He distinguishes characteristics from state quantities by their dynamics. State quantities change over the considered time scale and thereby parameterize the timewise behavior of systems while characteristics can be viewed as constant and therefore are well-suited to classify systems. We may add that in contrast to a state quantity, a characteristic like ''stability'' may not be possibly represented explicitly by the system at all. For a classification of system properties in this sense, see \cite{Reich2016_Quality}.  
 IEC61987 \cite{IEC61987} is an example of a characteristic-based catalog standard of classes of systems.

Ulrich Epple \cite{Epple2011_Merkmale} gives two examples for hierarchical relations. One for types of the carrier of the characteristics (that is, systems) on different levels of abstraction: a measuring device with the characteristic ''measurement range'' is more abstract than a flow meter with a ''cross section'' is more abstract than an inductive flow meter with a ''minimum conductivity''. This hierarchy fits nicely with the truncation/extension relation of data types. Especially as he demands that the less abstract device must ''inherit'' all characteristics of the more abstract device. The other hierarchy specializes characteristics: an inner diameter specializes an diameter specializes a length. Our usage of this example further above shows that this hierarchy fits nicely with the restriction/expansion of data types. 

In summary, the data concept with its data types and type hierarchies match the proposed structure of system characteristics.

%
\section{Discussion}
%
The presented data model is essentially a type concepts that combines alphabets and sets of (curried) operations: data is information which we know in principle how to process. Comparing our definition with the initial Merriam-Webster definition shows that we are pretty close to the colloquial meaning of data.

As already Alonso Church pointed out, operations themselves can be typed. However, typing of operations is more complex than typing of simple values and is beyond the scope of this article. An operation $op:X\rightarrow Y$ can be represented by a character ''op'' as an element of an alphabet $V$ in the sense of a name together with a function $h:V\times X \rightarrow Y$, mapping the name together with the input parameter of $op$ onto $op(x)$. Given $op$, the function $h$ is trivial. So, in principle, operations can be represented by their names which can be treated as characters. But intuitively, an operation type is defined by requiring certain properties of its operations and therefore does not change just because we introduce a new operation. So the essential question is how to define the set of operation names $V$. In the case of ordinary characters, it was a simple question of definition. In the case of operations, one could think that all that is required to process an operation is to know its domain and codomain. This would make the definition of $V$ indirect as the set of all names of operations which have a given domain and codomain. However, a restriction condition would relate to the behavior of these operations. For example, we could restrict the operations to only sine and cosine operations and all processing curriedops could rely on this assumption. In essence, with typing operations, the problem of behavioral subtyping as described by Barbara H. Liskov and Jeannette M. Wings \cite{LiskovWing1994} comes into the fore.   

There is no way to derive some canonical set of operations from an alphabet. We interpret the freedom to relate alphabets and sets of operations as the possibility to express our intent of the meaning of characters of the alphabet in an abstract sense. 
If we say that a certain alphabet should represent for example a temperature and not a velocity or something else, we determine that it can only be processed by operations that are intended to work on values of temperature. We therefore must know beforehand what a temperature is as far as the construction of the operations requires it.

It is interesting to see that simple type composition as an extension mechanism does not result, in general, in safe type relations. The main reason is the lack of a projection function. So, to type-safely extend a data type $country$ with the elements $\{England, France, Germany\}$ with a new country name $Spain$ requires that there is a sensible projection of the new element of $Spain$ to any of the old elements. This will usually require an element like $default$ or $unknown$ in the original alphabet. 

Please note, that a data type in the mentioned sense is a mathematical structure where the set of operations is not explicitly given. This is in contrast to abstract data types or objects in the object oriented sense whose sets of operations are usually comparatively small and, even more importantly, explicitly given. For example, the semantics of the C-data type ${\tt double}$ does not change if we add a new operation that is supposed to process a double variable, which would be the case for an abstract data type or an object.
However, there are some authors (e.g. Robert W. Sebesta, \cite{Sebesta2007}, p. 248) representing the idea that the set of operations of a type is predefined in the sense of objects.

With this conception of type semantics, the role and limitations of common type systems to facilitate interoperability becomes better comprehensible. Agreeing on common data types within an interaction implies that every interaction partner now has exactly the information she needs to avoid an unintended mismatch between the structure of the received information and the structural expectations of the operations with respect to their input. How much semantic connotation is provided by a type depends on how specific the concept is, it represents. 
However, as the nondeterministic interactions of networking, so called reactive systems cannot be represented by operations, mapping characters to characters (e.g. \cite{Reich2012_PRI}), there are principal limitations to this type semantics. 

It is obvious that our tools to create operations, namely modern imperative programming languages, should contain language elements to describe data types and their relations in the sense of this article. It is therefore quite surprising that virtually no modern programming language that we know of is expressive enough to represent the complete data type relation model of this article. It would be an endeavor of its own to investigate what aspects of our proposed type model can be found in which programming language. C allows the definition of composed types and also operation types but does not support any type relations. Pure so called ''object oriented language'' not even allow the declaration of data types, but only so called ''classes'', although classes without attached methods and only dynamic instance-related parameters could be viewed as data types in the sense of this article. Script languages like ECMAScript often are only very weakly typed. The language ADA is an example of a programming language that actually supports data type restrictions. For example {\tt subtype Int10 is Integer range 1..10;} defines Int10 as an integer type with a restricted value set of 1 \dots 10. The subranges of Pascal is a similar constructs. 

Currently we see a dramatic increase in the interest in data-oriented computing, like in the area of big data. We think that it is important to understand that ''data'' based on the concepts of information and types is to be understood not as a syntactic, but as a semantic concept that is directly related to the processing of the information. We think that it would be worthwhile to develop truly data oriented programming paradigms based on the presented type concept. Due to the much more flexible relation between alphabets and operations in the world of types compared to the world of objects, we would expect a data oriented programming paradigm also to be much more flexible.

\bibliographystyle{splncs}
\bibliography{bib/informatics}

\end{document}